\documentstyle[12pt]{article}

 \newcommand{\be}{\begin{equation}}
\newcommand{\ee}{\end{equation}}
\newcommand{\bea}{\begin{eqnarray}}
\newcommand{\eea}{\end{eqnarray}}

\input{amssymb.sty}

\newcommand{\PR}[1]{Phys. Rev.\ {\bf #1}\ }

\newcommand{\IJMP}[1]{Int. J. Mod. Phys.\ {\bf #1}\ }

\begin{document}
\begin{titlepage}
\begin{flushright}
{\large \bf UCL-IPT-97-18}
\end{flushright}
\begin{center}
\vskip 2cm

{\Large \bf
Isospin Amplitudes and CP violation in $(B \to K \pi)$ decays} \vskip
.4in

{\large 
J.-M. G\'erard and J. Weyers}\\[.15in] 

{\em Institut de Physique Th\'eorique\\
Universit\'e catholique de Louvain\\
B-1348 Louvain-la-Neuve, Belgium}

\end{center}

\vskip 2in

\begin{abstract}

We present a simple isospin invariant parametrization for $(B \to K\pi)$ decay amplitudes
which consistently includes CP violation and (quasi-elastic) hadronic final state
interactions. We find that the observed $(B \to K \pi)$ decays do not lead to a significant
bound on the angle $\gamma$ of the unitarity triangle. On the other hand, we claim that a
sizeable CP violation asymmetry in $(B^\pm \to K \pi^\pm)$ rates is by no means excluded.

\end{abstract}

\end{titlepage}

\newpage
\renewcommand{\thepage}{\arabic{page}}
\setcounter{page}{1}
\setcounter{footnote}{1}

\section{Introduction.}

The Standard Model (SM) encodes a very neat parametrization of quark mixing and CP
violation through the Cabibbo-Kobayashi-Maskawa matrix (CKM matrix). At present, all
existing data are consistent with this parametrization although precise experimental tests of
the pattern of CP violation are still lacking. With $B$-factories forthcoming, the situation will
soon improve and, at least in principle, detailed checks of the SM predictions for CP
violation will become possible.

With this exciting perspective in mind, a huge amount of work has been devoted in recent
years to possible ways of extracting information on CP violation from various $B$ decays
\cite{1}. The difficulty is of course what to do with the ``hadronic complications" which are
unavoidable when physical parameters, precisely defined at the quark level, have to be
related to measurable quantities in the hadronic world.

In this note, we reconsider the $(B \to K \pi)$ decay amplitudes and advocate the use of a
hadronic basis, namely the {\it isospin} basis. This approach provides a simple bookkeeping
procedure for (quasi-elastic) hadronic final state interaction effects. As a direct result we
find that the observed $(B \to K \pi)$ decays do not lead to a significant bound on the
$\gamma$ angle of the unitarity CKM triangle without specific unwarranted assumptions
on ``hadronic effects". On the other hand, we emphasize that the CP asymmetry in $(B
^\pm \to K \pi^\pm)$ can be quite large.

\section{Isospin amplitudes.}

Isospin is a good symmetry of the hadronic world! In lowest order the SM weak Hamiltonian
responsible for the $(B \to K \pi)$ decays contains both an isosinglet $(H^o_W)$ and an
isotriplet  $(H^1_W)$ part. The $B$ mesons $(B^+,B^o)$ are of course an isodoublet while
the $(K \pi)$ system is a mixture of $I = 1/2$ and $I=3/2$ eigenstates.

Let
$$
a_1 \equiv \langle \langle B | H^o_W | (K \pi), I = 1/2 \rangle \rangle
\eqno{(1.a)}
$$
$$
b_1 \equiv \langle \langle B | H^1_W | (K \pi), I = 1/2 \rangle \rangle
\eqno{(1.b)}
$$
$$
b_3 \equiv \langle \langle B | H^1_W | (K \pi), I = 3/2 \rangle \rangle
\eqno{(1.c)}
$$
be the reduced matrix elements of the weak Hamiltonian. Let us also, for simplicity,
consider the quasi-elastic approximation for the $(K \pi)$ system: all final state interaction
effects are described by $\delta_1$ and $\delta_3$, the $s$-wave phase shifts in the $I =
1/2$ and $I = 3/2$ channels.

From all these old fashioned trivialities, one readily obtains
$$
A (B^+ \to K^o \pi^+) = \sqrt{\frac{2}{3}} a_1 e^{i \delta_1} + \frac{\sqrt{2}}{3} b_1 e^{i
\delta_1} - \frac{\sqrt{2}}{3} b_3 e^{i \delta_3}
\eqno{(2.a)}
$$
$$
A (B^o \to K^+ \pi^-) = \sqrt{\frac{2}{3}} a_1 e^{i \delta_1} - \frac{\sqrt{2}}{3} b_1 e^{i \delta_1}
+ \frac{\sqrt{2}}{3} b_3 e^{i \delta_3}
\eqno{(2.b)}
$$
and similar expressions for the other channels.

\section{Quark diagrams.}

The isospin invariant amplitudes $a_1$ and $b_{1,3}$ receive contributions from various
SM quark diagrams. It is only at the level of these diagrams that the specific CP violation
pattern of the SM can be correctly implemented. On the other hand all QCD effects are
isospin invariant and can thus be ignored for our purposes.

For simplicity, let us only keep the contributions to $(a_1,b_1,b_3)$ coming from the
so-called \cite{1} tree-level $(T)$, color-suppressed $(C)$ and QCD-penguin $(P)$ quark
diagrams.

Thus we write
\be
\setcounter{equation}{3}
a_1 = a^T_1 e^{i\gamma} + a^C_1 e^{i\gamma} + a^P_1 + \cdots
\ee
and similar expressions for $b_1$ and $b_3$. In Eq.(3), $\gamma$ is of course the
CP-violating phase coming from the $V^\ast_{ub}$ CKM matrix element while the dots
stand for neglected contributions such as the annihilation amplitude.

It is now straightforward to derive the relations
$$
a^T_1 = -\sqrt{3} b^T_1 \ \ \ , \ \ \ b^T_3 =  - 2 b^T_1
\eqno{(4.a)}
$$
$$
a^C_1 = 0 \ \ \ \ \ \ \ \ \ \  \  , \ \ \  b^C_3 =  b^C_1 \ \ \ \ 
\eqno{(4.b)}
$$
$$
b^P_1 = 0 \ \ \  \ \ \ \ \ \  \ \ , \ \ \ b^P_3 =  0 . \ \ \ \ 
\eqno{(4.c)}
$$
We redefine
$$
T \equiv -2\sqrt{2} b^T_1 \eqno{(5.a)}
$$
$$
C \equiv \sqrt{2} b^C_1 \ \ \ \  \eqno{(5.b)} 
$$
$$
P \equiv \sqrt{\frac{2}{3}} a^P_1 \ \ \ \ \eqno{(5.c)}
$$
and multiply Eqs.(2) by an overall phase $e^{-i\delta_1}$ to obtain finally
$$
\tilde A (B^+ \to K^o \pi^+) = \frac{1}{3} (1-e^{i\delta}) (T+C) e^{i\gamma} + P \ \ \ \ \ \   \ \ \ \ \
\ \ \ 
\eqno{(6.a)}
$$
$$
\tilde A (B^o \to K^+ \pi^-) = \frac{1}{3} (1-e^{i\delta}) (2T-C) e^{i\gamma} + T e^{i\gamma}
e^{i\delta} + P
\eqno{(6.b)}
$$
where $\delta = \delta_3 -  \delta_1$.

Eqs.(6) consistently include both CP violation as prescribed in the SM and final state
interactions as constrained by isospin invariance.

Note in particular that Eqs.(6) are  {\it not} equivalent to a commonly used \cite{1} quark
parametrization where $T$ and $P$ are given strong phases $\delta_T$ and $\delta_P$,
respectively. The latter parametrization is not compatible \cite{2} with isospin invariance
unless
$\delta = \delta_3 - \delta_1 = \delta_T - \delta_P = 0 $ !

\section{Comments and applications.}

The presence of a color-suppressed amplitude $C$ in Eqs.(6) confirms that the
``quasi-elastic" rescatterings
$$
B^+ \to \{ K^+ \pi^o \} \to K^o \pi^+ 
\eqno{(7.a)}
$$
$$
B^o \to \{ K^o \pi^o \} \to K^+ \pi^-
\eqno{(7.b)}
$$
are correctly included in our formalism. So, we do not have to invoke penguin
topology \cite{3} with internal up-quark exchange: here $P$ is a real amplitude while
$e^{i\delta}$ and $(1-e^{i\delta})$ consistently approximate in an isospin invariant
way final state interactions.

Color-allowed penguin amplitudes $P_{EW}$ are second order weak effects. At this order,
the weak Hamiltonian acquires, in general, an extra $I=2$ piece, $H^2_W$, with reduced
matrix element
\be
\setcounter{equation}{8}
c_3 \equiv \langle \langle B | H^2_W | (K \pi) , I = 3/2 \rangle \rangle  .
\ee

However, the dominant electroweak diagrams (with a top intermediate state) have only
$I=0$ and $I=1$ pieces and are thus easily included in Eqs.(6) via the substitution
\be
C e^{i\gamma} \mapsto C e^{i\gamma} + P_{EW}  .
\ee

In the applications to follow, let us however neglect these potentially large contributions.

\subsection{The Fleischer-Mannel (FM) bound.}

In the approximations made, we now define the (real) ratio
\be
r = \frac{T}{P}
\ee
and consider
\be
R \equiv \frac{\Gamma (B^o \to K^+ \pi^-) + \Gamma (\bar B^o \to K^- \pi^+)}{\Gamma
(B^+ \to K^o \pi^+) + \Gamma (B^- \to \bar K^o \pi^-)}
\ee
recently measured by the CLEO Collaboration \cite{4} to be $R = 0.65 \pm 0.40$.

From Eqs.(6) one easily obtains the constraint
\be
\sin^2 \gamma \leq 1 - \frac{(1-R)[5-2R+2(2+R)\cos \delta]}{[2-R+(1+R)\cos \delta]^2}
\ee
Obviously for $\delta = 0$, but only in this case, one recovers the FM bound \cite{5}, namely
$\sin^2 \gamma \leq R$.  Clearly this latter bound is in general not
valid and the constraint on $\gamma$ given in Eq.(12) does depend on hadronic physics
via the free parameter $\delta$.

In other words, even if $R$ turns out to be strictly less than unity, this does by no means
exclude any value of  the angle $\gamma$ of the unitarity triangle, including $\gamma =
\pi/2 $ !

\subsection{CP asymmetry in $B^\pm \to K \pi^\pm$.}

Again a simple calculation gives
$$
a \equiv \frac{\Gamma (B^+ \to K^o \pi^+) - \Gamma (B^- \to \bar K^o \pi^-)}{\Gamma
(B^+ \to K^o \pi^+) + \Gamma (B^-  \to  \bar K^o \pi^-)}
$$
\be
\simeq \frac{2}{3} \  r \  \sin \gamma \sin \delta \ \ \ \ \ \ \ \ \   \ \ \ \ \ \ \ \ \   \ \ \ \ \ \ 
\ee
if QCD-penguin dominates this $B \to K \pi$ channel (i.e. $r < 1$). Clearly this asymmetry
could be very sizeable contrary to recent claims \cite{6} based  on a quark
parametrization with
$\delta_{T,P}$.

Needless to say,  our conclusions on the FM bound and on the $B^\pm \to K^o \pi^\pm$
asymmetry are reinforced if one takes into account color-allowed electroweak penguin
contributions. The estimates \cite{1}
\be
| \frac{T}{P} | = {\cal O}  (0.2) \ \ \ \ \ , \ \ \  | \frac{P_{EW}}{T} | = {\cal O} (1)
\ee
based on factorization and $SU(3)$ relations suggest that $T$ and $P_{EW}$ amplitudes are
equally important. In such a case, the bound on $\sin^2 \gamma$ becomes totally useless
while the $B^\pm \to K \pi^\pm$ CP asymmetry can be as large as 10$\%$.

\section{Conclusion.}

The quasi-elastic approximation made in this note has the main advantage of being
simple. Clearly a more sophisticated analysis of the final state interactions is possible but it
will not change our main point: the strong phases associated with $a_1$ and $b_1$ will be
equal and will differ from the strong phase for $b_3$ (isospin invariance!).

It is straightforward to extend our analysis to more and more approximate flavor
symmetries. In particular, a possible bonus is that rescattering effects of the type 
$\{ K \pi \} \rightleftharpoons \{ \bar D D_s \}$
 can then be treated as quasi-elastic within
$SU(4)_f$.

To conclude, let us stress once again the main point of this note: isospin is an excellent
symmetry of the strong interactions and parametrizations of various mesonic decay
amplitudes should at least be compatible with it !

\vspace*{1cm}
\noindent
{\large\bf Acknowledgements}\\

We are grateful to Jean Pestieau for  useful discussions and comments.

\end{document}